# Representing Digital Assets for Long-Term Preservation using MPEG-21 DID

Jeroen Bekaert[1,2], Xiaoming Liu[1], and Herbert Van de Sompel[1]

[1] Digital Library Research & Prototyping Team, Los Alamos National Laboratory, MS P362, PO Box 1663, Los Alamos, NM 87544-7113, US
{jbekaert, liu_x, herbertv}@lanl.gov
[2] Dept. of Architecture and Urbanism, Faculty of Engineering, Ghent University, Jozef-Plateaustraat 22, 9000 Gent, Belgium
{jeroen.bekaert}@ugent.be

**Abstract.** Various efforts aimed at representing digital assets have emerged from several communities over the last years, including the Metadata Encoding and Transmission Standard (METS), the IMS Content Packaging (IMS-CP) XML Binding and the XML Formatted Data Units (XFDU). The MPEG-21 Digital Item Declaration (MPEG-21 DID) is another approach that can be used for the representation of digital assets in XML. This paper will explore the potential of the MPEG-21 DID in a Digital Preservation context, by looking at the core building blocks of the OAIS Information Model and the way in which they map to the MPEG-21 DID abstract model and the MPEG-21 DIDL XML syntax.

## 1 Introduction

Digital assets stored in archival systems are typically compound, consisting of multiple datastreams of a variety of MIME media types. They also hold secondary information, such as information supporting discovery, digital preservation and rights management. The ISO Open Archival Information System (OAIS) Reference Model [1] refers to these digital assets as Content Information; the Kahn/Wilensky framework [2] refers to them as digital objects.

Various efforts aimed at representing such digital assets have emerged from several communities over the last years [3]. Of special interest in the current technological environment are techniques that represent a digital asset as an XML document. Examples include, the Metadata Encoding and Transmission Standard (METS) [4], an initiative of the Digital Library Federation; the IMS Content Packaging (IMS-CP) XML Binding [5], a representational technique predominantly used in the educational domain; and the XML Formatted Data Units (XFDU) [6], a Standard under development by the Consultative Committee for Space Data Systems (CCSDS) aimed at aligning a representational approach with the OAIS Reference Model [1].

MPEG-21 Digital Item Declaration (MPEG-21 DID) [7,8] is another specification that can be used for the representation of digital assets in XML. MPEG-21 DID is the second part of the MPEG-21 suite of ISO Standards, under development by the Moving Picture Experts Group (MPEG). The MPEG-21 suite currently consists of eighteen parts, each of which can be used separately, but is developed to give optimal result when used together. For example, the combination of MPEG-21 DID – which focuses on the representation of digital assets – with MPEG-21 Digital Item Identification (MPEG-21 DII) [9] results in an unambiguous approach to handle identifiers of digital assets. The fifth part of MPEG-21, the Rights Expression Language [10], details a language to express rights pertaining to digital assets. And the ninth part of MPEG-21, the MPEG-21 File Format [11], specifies a technique to serialize a digital asset represented according to MPEG-21 DID in a single binary package.

Although MPEG-21 originates in a community that mainly focuses on the coding of audio and video, MPEG-21 DID can be used to represent any kind of compound digital asset, including electronic journals, scientific datasets, e-learning objects, broadcasting data, and so on. MPEG-21 DID takes an approach that enforces cross-community interoperability, while allowing the flexibility for the emergence of compliant, domain-specific profiles. One such profile could focus on the representation of digital assets for long-term preservation. This paper shows that MPEG-21 DID can also be aligned with the OAIS Reference Model [1].

## 2 MPEG-21 Digital Item Declaration

The MPEG-21 Digital Item Declaration (MPEG-21 DID) [7,8,12,] specification consists of 2 parts. It starts by introducing a set of abstract entities that, together, form a well-defined data model for declaring digital assets. In the abstract model, the entity that represents a digital asset is referred to as a *Digital Item*. Next, based on the entities defined in the abstract model, MPEG-21 DID defines a serialization of the model in XML. This XML syntax, formalized by means of an XML Schema [13], is called the MPEG-21 Digital Item Declaration Language (MPEG-21 DIDL). In MPEG-21 DIDL, a *Digital Item* is represented by an `Item` XML element from the MPEG-21 DIDL XML Namespace. The XML document containing the XML declaration of the *Digital Item* is called a DIDL document; it starts with a `DIDL` root element.

In this paper, the `courier` font is used to refer to XML elements of the MPEG-21 DIDL; the *italic* font is used when a reference is made to entities of the MPEG-21 DID abstract data model.

A first edition of MPEG-21 DID was published as an ISO Standard in March 2003 [8]. A second edition [7] has been finalized mid 2005 and mainly augments the functionality of the MPEG-21 DIDL. Once published, the second edition of the MPEG-21 DID specification will be freely available from the ISO website [http://www.iso.org]. Interested readers are referred to [12] for more details on the use of the MPEG-21 DID abstract data model and its MPEG-21 DIDL XML serialization.

## 3 MPEG-21 DID and the OAIS Reference Model

The Reference Model for an Open Archival Information System (OAIS) developed by the CCSDS has become a foundation for thinking about long-term preservation problems. The OAIS model defines both a Functional Model and an Information Model. The Functional Model outlines the range of functions that need to be undertaken by a compliant archive, such as access, archival storage, and ingest. The Information Model defines broad types of information that are required in order to preserve and access the information stored in an archive. The core building of the OAIS Information Model – and the way in which they map to the MPEG-21 DID abstract model and the MPEG-21 DIDL XML syntax – are depicted in Figure 1 and explained below:

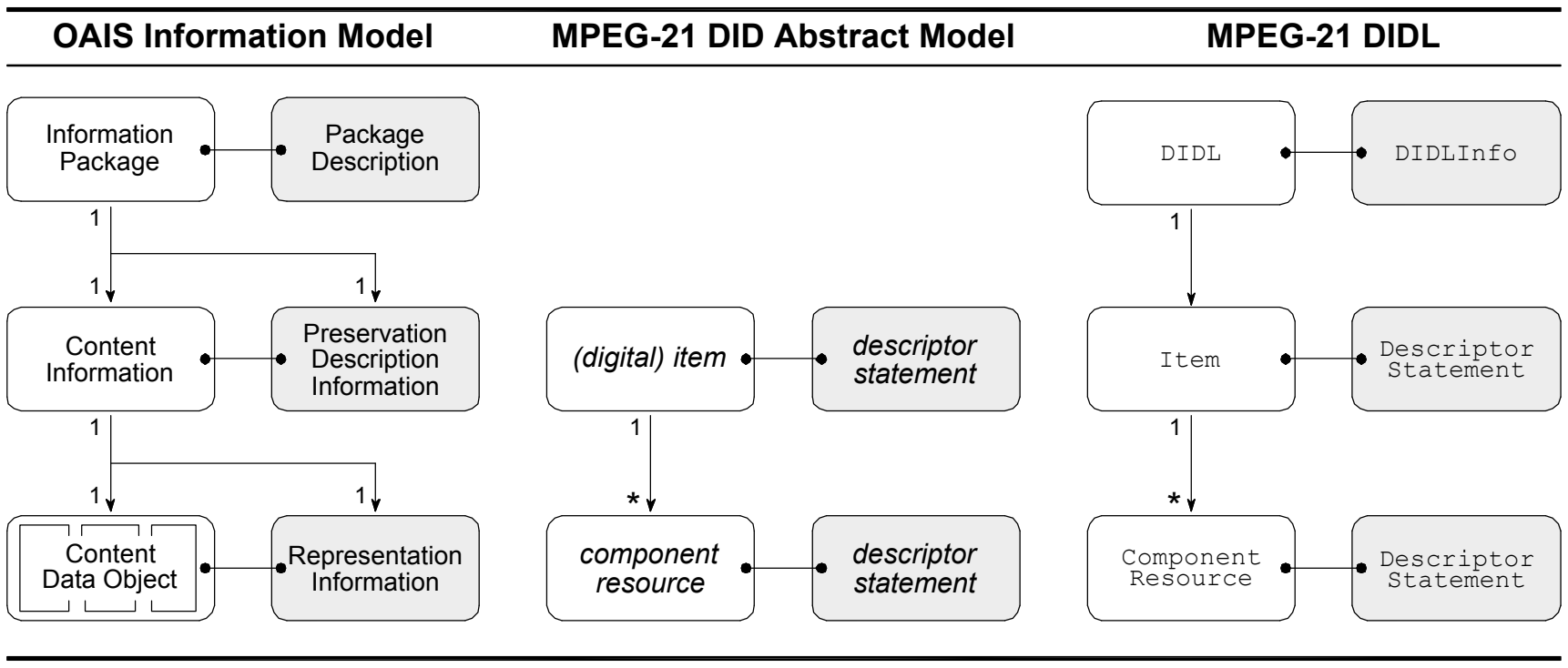

**Fig. 1.** Mapping concepts from the OAIS Information Model to the MPEG-21 DID abstract model and the MPEG-21 DIDL XML syntax

In the OAIS Information Model one may approximately equate a *Digital Item* with the concept of Content Information. Indeed, Content Information is considered the original target of preservation, and hence the main point of focus, in an OAIS environment. It may be important to note that MPEG-21 DID allows for a *Digital Item* to be composed of other *Digital Items*. The OAIS Information Model, however, does not explicitly recognizes the notion of Content Information being composed of other Content Information. Because of this, and for reasons of clarity, the MPEG-21 DID *item* nesting capabilities are not used in the described mapping.

According to the OAIS, Content Information is comprised of a Content Data Object combined with secondary information related to the Representation of the Content Data Object. A Content Data Object is typically a sequence of bits that can be implemented as one or many files. In MPEG-21 DID, each of those files is represented by a *component/resource* construct (see Section 3.1). Representation Information pertaining to the Content Data Object can be conveyed using *descriptor/statement* constructs at the *component*-level (see Section 3.2).

In addition, the OAIS Information Model defines Preservation Description Information to adequately preserve the particular Content Information with which it is associated. The Preservation Description Information can be further broken down into four sub-categories: Reference Information, Context Information, Provenance Information, and Fixity Information. Again, in MPEG-21 DID, Preservation Description Information can be provided using *descriptor/statement* constructs (see Section 3.2). These *descriptor/statement* are appended to the *item*-level, providing secondary information about the *Digital Item*.

Finally, the OAIS Information Model defines Packaging Information as the information structure that binds and encapsulates the Content Information and Preservation Description Information into an identifiable Information Package. In MPEG-21 DIDL, this package is represented in XML and is referred to as the DIDL document. Also according to the OAIS Information Model, Package Descriptions may be added to the Information Package to contain secondary information about the Information Package itself. MPEG-21 DIDL provides the flexibility to convey secondary information about a DIDL document using `DIDLInfo` elements (see Section 3.3).

In this paper, references will be made to a sample set of Content Information to illustrate the main characteristics of MPEG-21 DIDL in the context of its use in the digital preservation domain. The sample Content Information is a photograph with identifier 'urn:foo/015997845'. The Content Information consists of a Content Data Object which, in turn, is made up of 2 files: a TIFF image file, and a JPEG2000 image file. The DIDL document representing the Content Information is shown in full in Appendix A.

### 3.1 Providing a Content Data Object

A Content Data Object, as specified by the Information Model for an OAIS, can be mapped to one or more *component/resource* constructs of the MPEG-21 DID abstract data model, depending on whether a Data Object is implemented using one or more files. In MPEG-21 DIDL, *component* and *resource* entities are represented by a `Component` and `Resource` XML element, respectively. MPEG-21 DIDL provides the ability to include *resources* in two ways: a *resource* can be embedded in a DIDL document using a base64 encoding algorithm [14] – By Value – or can be pointed at from within a DIDL document – By Reference. Also, MPEG-21 DIDL explicitly supports the provision of multiple bit equivalent files (by grouping those files in a single `Component` element), and provides a mechanism to deal with compressed files that both safeguards the original MIME media (sub)type and provides information on the compression algorithms being used.

The DIDL document in Appendix A shows a By Reference provision of both the TIFF and JPEG2000 files pertaining to the Content Data Object of the sample Content Information. A `ref` attribute of the `Resource` element contains the actual network location. A mandatory `mimeType` attribute indicates the MIME media and subtype of the files.

### 3.2 Providing Representation and Preservation Description Information

MPEG-21 DID offers a flexible, and extensible approach to convey secondary information pertaining to entities of the abstract data model, using so-called *descriptor/statement* constructs. In MPEG-21 DIDL, these *descriptor/statement* constructs are represented using `Descriptor` and `Statement` elements. The MPEG-21 Standard itself defines ways to use these constructs as a means to convey – amongst others – identification information (as specified by MPEG-21 DII) [9], and rights information (as specified by MPEG-21 REL) [10]. But the same constructs can be used to convey domain, or application specific secondary information. As a result, secondary information of particular importance for the preservation of Content Information can also be conveyed using *descriptor/statement* constructs.

MPEG-21 DID does not provide any categorization nor presupposition about the types of secondary information that can be conveyed using *descriptor/statement* constructs. While METS and XFDU impose some kind of categorization by pre-defining placeholders for descriptive, adminstrative, rights, source and provenance information, MPEG-21 DIDL relies on the XML Namespaces of the information provided in `Descriptor/Statement` constructs to determine the semantics of the conveyed information. For example, according to MPEG-21 DII, XML elements in the MPEG-21 DII XML Namespace provide information that allows for the identification of a Digital Item or part thereof. This approach also allows for application domains to profile DIDL documents by restricting the XML Namespaces that can be used, and by specifying the semantics of the use of elements of those Namespaces. As shown in 3.2.1 and 3.2.2, this approach can also be used is a digital preservation context.

#### 3.2.1 Representation Information

`Descriptor/Statement` constructs at the *component*-level can be used to provide Representation Information. Appendix A shows the use of two `Descriptor/Statement` constructs to provide format-specific representation information pertaining to the TIFF and JPEG2000 constituents of the Content Data Object, respectively. To that end, XML-based information about the constituents, generated by the JSTOR/Harvard Object Validation Environment (JHOVE) [15], is encapsulated in a `Descriptor/Statement` construct. The `Descriptor/Statement` construct and the `Resource` element (providing the TIFF or JPEG2000 files) are bound by a `Component` element.

#### 3.2.2 Preservation Description Information

Similarly, `Descriptor/Statement` constructs at the *(digital) item*-level can be introduced that allow expressing Preservation Description Information. For example:

- A `Descriptor/Statement` construct can be added to an `Item` element, to provide the identifier (i.e. Reference Information) of the Content Information. Appendix A shows the use of a `Descriptor/Statement` construct containing the `Identifier` element from the MPEG-21 DII XML Namespace to associate the URI 'urn:foo/015997845' with the sample Content Information.
- To allow verifying the integrity and authenticity of a Content Data Object (and/or its constitutents), W3C Signature XML Signatures [16] can be provided in `Descriptor/Statement` constructs attached to the `Component` elements that contains the constituents of the Content Data Object as a `resource`. In the OAIS Information Model, this type of information is categorized as Fixity Information.
- Similarly, other Preservation Description Information can be added as required. For example, Information related to the digital preservation of an asset can also be conveyed by elements from the PREMIS effort [http://www.oclc.org/research/projects/pmwg/]. Again, such information can be provided in `Descriptor/Statement` constructs.

### 3.4 Providing Package Descriptions

The second edition of MPEG-21 DID allows for the association of secondary information pertaining to the DIDL document itself by adding attributes to the `DIDL` root element or by including XML content in the `DIDLInfo` element that itself is a child of the `DIDL` root element. Also, the `DIDL` root element may have an optional `DIDLDocumentId` attribute. This attribute can be used to convey the identifier of the DIDL document. In the OAIS Information Model, the latter is referred to as the Information Package Identifier.

The full example in Appendix A shows the use of the `DIDLInfo` element to convey a W3C XML Signature [16] calculated over the DIDL document itself. This XML Signature allows checking the integrity of the Information Package stored in an OAIS. The example also shows the use of the `DIDLDocumentId` attribute to convey the Information ackage Identifier.

## 4 Conclusion

This paper has shown that the MPEG-21 DID abstract model and MPEG-21 DIDL can be aligned with concepts of the OAIS Information Model. The use of the MPEG-21 DID Standard for the representation of digital assets in Digital Preservation applications is attractive. From a strategic perspective, the appeal stems from MPEG-21 DID being part of the ISO MPEG-21 suite of Standards, and hence is likely to receive strong industry backing. MPEG-21 DID is developed by major players in the content and technology industry, which provides some guarantees regarding its adoption and, hence regarding the future emergence of compliant migration tools. Also, MPEG-21 DID is appealing because its abstract data model facilitates the long-term use of the same conceptual model, which can be instantiated in different serializations (the XML-based DIDL serialization only being one) as technologies evolve.

From a functional perspective, the MPEG-21 DIDL XML syntax is attractive because:

- MPEG-21 DIDL provides the ability to include constituents of a Content Data Object in two ways: a constituent can be contained in a DIDL document – By Value – or can be pointed at from within a DIDL document – By Reference.
- The `Descriptor/Statement` constructs of MPEG-21 DIDL offer a highly flexible, extensible approach to convey secondary information pertaining to the Content Information and Content Data Object. As was shown, secondary information of particular importance in a preservation context can readily be included. Examples include Representation Information (e.g. format-specific information generated by JHOVE), Reference Information (e.g. MPEG-21 DII) Preservation metadata (e.g. PREMIS) and Fixity Information (e.g. W3C XML Signatures).
- The `DIDLInfo` element of MPEG-21 DIDL provides the same flexibility to convey secondary information about an Information Package (a DIDL document), as the `Descriptor/Statement` constructs provide for the Content Information and Content Data Object.

## Appendix A: Sample DIDL document

```
<?xml version="1.0" encoding="UTF-8"?>
<didl:DIDL DIDLDocumentId="info:lanl-repo/i/11d8-a819-b1db893d21e6"
  xmlns:didl="urn:mpeg:mpeg21:2002:02-DIDL-NS">
  <didl:DIDLInfo>
    <dsig:Signature
      xmlns:dsig="http://www.w3.org/2000/09/xmldsig#">
     <!--  W3C XML Signature of Information package -->
    </dsig:Signature>
  </didl:DIDLInfo>
  <didl:Item id="uuid-00004342-c477-11d8-a819-b1db893d21e6">
    <didl:Descriptor>
      <didl:Statement mimeType="application/xml; charset=utf-8">
        <dii:Identifier xmlns:dii="urn:mpeg:mpeg21:2002:01-DII-NS">
          urn:foo/015997845</dii:Identifier>
      </didl:Statement>
    </didl:Descriptor>
    <didl:Component id="uuid-00005e90-c687-11d8-a819-b1db893d21e6">
      <didl:Descriptor>
        <didl:Statement mimeType="application/xml; charset=utf-8">
          <jh:jhove
            xmlns:jh="http://hul.harvard.edu/ois/xml/ns/jhove">
            <!--  JHOVE Information for Content Data Object -->
          </jh:jhove>
        </didl:Statement>
      </didl:Descriptor>
      <didl:Descriptor>
        <didl:Statement mimeType="application/xml; charset=utf-8">
          <dsig:Signature
            xmlns:dsig="http://www.w3.org/2000/09/xmldsig#">
            <!--  W3C XML Signature of Content Data Object -->
          </dsig:Signature>
        </didl:Statement>
      </didl:Descriptor>
```

```
      <didl:Resource mimeType="image/tiff"
         ref="http://foo/bar/pict/015997845.tiff"/>
    </didl:Component>
    <didl:Component id="uuid-0000a01c-d579-21d8-a819-b1db893d21e6">
      <didl:Descriptor>
        <didl:Statement mimeType="application/xml; charset=utf-8">
          <jh:jhove
            xmlns:jh="http://hul.harvard.edu/ois/xml/ns/jhove">
            <!--  JHOVE Information of Content Data Object -->
          </jh:jhove>
        </didl:Statement>
      </didl:Descriptor>
      <didl:Descriptor>
        <didl:Statement mimeType="application/xml; charset=utf-8">
          <dsig:Signature
            xmlns:dsig="http://www.w3.org/2000/09/xmldsig#">
            <!--  W3C XML Signature of Content Data Object -->
          </dsig:Signature>
        </didl:Statement>
      </didl:Descriptor>
      <didl:Resource mimeType="image/jp2"
         ref="http://foo/bar/pict/015997845.jp2"/>
    </didl:Component>
  </didl:Item>
</didl:DIDL>
```